\documentclass[aps,prd,twocolumn,groupedaddress]{revtex4-1}
\usepackage{amsmath}
\usepackage{graphicx}
\usepackage{dcolumn}
\begin{document}

%\preprint{}

\title{Neutrinos and dark energy constraints from future galaxy surveys and CMB lensing information}


\author{L. Santos}
\email{larissa.santos@roma2.infn.it}
%\thanks{}
\affiliation{Dipartimento di Fisica, Universit\`a di Roma ``Tor Vergata''}
\author{P. Cabella}
\affiliation{Dipartimento di Fisica, Universit\`a di Roma ``Tor Vergata''}
\author{A. Balbi and N. Vittorio}
\affiliation{Dipartimento di Fisica, Universit\`a di Roma ``Tor Vergata"}
\affiliation{INFN, Sezione di Roma Tor Vergata}


\date{\today}

\begin{abstract}

 We explore the possibility of obtaining better constraints  from future astronomical data by means of the Fisher information matrix formalism. In particular, we consider how cosmic microwave background (CMB) lensing information can improve our parameter error estimation.  We consider a massive neutrino scenario and a time-evolving dark energy equation of state in the $\Lambda$CDM framework. We use Planck satellite experimental specifications together with the future galaxy survey Euclid in our forecast. We found improvements in almost all studied parameters considering Planck alone when CMB lensing information is used. In this case, the improvement with respect to the constraints found without using CMB lensing is of  93\% around the fiducial value for the neutrino parameter.  The improvement on one of the dark energy parameter reaches 4.4\%. When Euclid information is included in the analysis, the improvements on the neutrino parameter constraint is of approximately 128\%  around its fiducial value. The addition of Euclid information provides smaller errors on the dark energy parameters as well. For Euclid alone, the FoM is a factor of $\sim$ 29 higher than that from Planck alone even considering CMB lensing. Finally, the consideration of a nearly perfect CMB experiment showed that CMB lensing cannot be neglected specially in more precise future CMB experiments since it provided in our case a 6 times better FoM in respect to the unlensed CMB analysis .

\end{abstract}

%\keywords{}

\maketitle

\section{Introduction\label{Introduction}}

The discovery of the accelerated expansion of the universe \citep{riess, per} can be interpreted by introducing in the cosmological model a negative pressure component, termed "dark energy".  The simplest dark energy candidate is a cosmological constant $\Lambda$, having constant equation of state $w_{de}= P_{de} / \rho_{de}= -1$. Together with a pressureless cold dark matter component (CDM), this constitutes the standard $\Lambda$CDM model.  Although this "concordance model" is in very good agreement with a variety of cosmological observations \citep{2011larson, 2011komatsu}, different candidates of dark energy cannot be discarded yet. Moreover, the basic cosmological constant scenario has two difficulties known as "fine-tuning" and cosmic coincidence problems (see, e.g., \cite{1999zlatev}). To overcome these problems alternative candidates for the dark energy have been proposed, such as the quintessence \citep{1998caldwell} that allows the possibility of a time-dependent equation of state \citep{2003linder}. In this paper, we will assume a redshift dependent equation of state for the dark energy,

\begin{equation}
\label {w_z}
w_{de} (z)= \frac{P_{de}(z) }{ \rho_{de}(z)}.
\end{equation}

and adopt the well-known Chevalier-Plarsky-Linder parametrization \citep{2001polarski, 2003linder}

\begin{equation}
\label {w_a}
w_{de}(a)= w_{0}+ (1-a)w_a.
\end{equation}

Cosmological observation can in principle be used to constrain the neutrinos' masses. It was shown by neutrino oscillation experiments that neutrinos have non-zero masses (see \citep{2012gonzales} and references therein). However, these experiments can only constrain the neutrinos mass-square differences and not their individual values (for a review in neutrino masses see \citet{2009degouvea}).  On the other hand, cosmological probes are most sensitive to the total neutrino masses, $\sum m_\nu$. Using CMB radiation data only, from Planck satellite,  an upper limit  to  $\sum m_\nu$ of 0.933 eV at 95\% C.L was found \citep{2013planckc}.

\begin{equation}
\label {omega_nu}
\Omega_{\nu}= \frac{m_{\nu}}{ 94h^2eV}.
\end{equation}

However, the dark energy equation of state and the neutrinos' total mass  parameters are degenerated (see, e.g.,\citep{2005hannestad}).  Some work has already been done to constrain both parameters simultaneously in a few dark energy scenarios, such as for models with a constant and a time-varying equations of state \citep{2008ichikawa, 2010hamann, 2011carboneb, 2012joudaki}.

Our goal is to forecast the constraint in total mass of neutrinos in a time evolving dark energy model, using  CMB temperature and polarization power spectrum from the Planck satellite experimental setup (including also CMB lensing information), as well as the large-scale matter distribution that can be observed by Euclid survey. We emphasize the usage of Planck CMB polarization information since its temperature data has been recently released \citep{2013planckc}.  We assume a geometrically flat $\Lambda$CDM model  with two massive neutrinos with identical mass and one massless in a inverted hierarchy mass splitting, being  $m_{\nu}$ = 0.125 eV for each massive neutrino.  The fiducial parameters are $h^2\omega_b=0.02219$, $h^2\omega_c=0.1122$, $h^2\omega_{\nu}=0.0027$, $h=0.65$, $n_s=0.952$. We normalize the CMB power spectra to COBE.  For similar approaches see \citet{2009hollenstein, 2010namikawa, 2012das, 2012hall, 2012hamann, 2012joudaki} . The paper is organized as follows: We give a small introduction on CMB lensing in Sec. \ref{CMB lensing}. In Sec. \ref{Method},  we briefly review the Fisher information matrix formalism for the CMB (with and without lensing information) and for a galaxy survey. Finally present our results in Sec. \ref{Results}, followed by our discussion and conclusions in Sec. \ref{Discussion and conclusions}.

%. We describe the fiducial model used in the analysis in Sec. \ref{Fiducial models} and

\section{CMB lensing\label{CMB lensing}}

A small effect that can be observed in CMB power spectrum regards the deflection of photons, during their travel between the last scattering surface and the observer, by gravitational potentials $\Psi$ due to clusters of galaxies.  \citet{2007smith} detected the CMB lensing signal for the first time by cross correlating WMAP data to radio galaxy counts in the NRAO VLA sky survey (NVSS). Recently, the detection of the gravitational lensing using CMB temperature maps alone and the measurement of the power spectrum of the projected gravitational potential were done using the Atacama Cosmology Telescope and the South Pole Telescope \citep{2011das, 2012engelen}.  

The lensing potential is defined as:

\begin{equation}
\label {pot_lens}
\psi(\textbf{\^n}) \equiv -2\int_{0}^{\chi^*}d\chi \frac{\chi^*- \chi} {\chi^*\chi} \Psi(\chi\textbf{\^n}; \eta_{0}-\chi),
\end{equation}
being $\chi^*$ the comoving distance and $\eta_{0}-\chi$ is the conformal time at which the photon was at position $\chi\textbf{\^n}$.

The lensing effect remaps the temperature and polarization fields as

\begin{equation}
\label {temp_lens}
\frac{\Delta\textrm{\~T(\textbf{\^n}})} {T} = \frac{\Delta\textrm{T(\textbf{\^n'})}}{T} = \frac{\Delta\textrm{T(\textbf{\^n}} + d)}{T},
\end{equation}

\begin{equation}
\label {pol_lens}
[Q + iU] (\textbf{\^n})  = [Q+ iU] (\textbf{\^n}+ d).
\end{equation}

where in the case of the temperature field, the temperature T of the lensed CMB in a direction \textbf{\^n} is equal to the unlensed CMB in a different direction \textbf{\^n'}.  Both these directions, \textbf{\^n} and \textbf{\^n'}, differ by the deflection angle $d$.  To first order, the deflection angle is simply the lensing potential gradient, $d= \nabla \psi$. In the same way,  the effect of lensing in CMB polarization is written in terms of the Stokes parameters $Q(\textbf{\^n})$ and $U(\textbf{\^n})$ (for a review in CMB polarization theory, see \citet{2004cabella}) :

CMB lensing has important quantitative contributions that should be taken into account, therefore a lot of work has been done to CMB lensing reconstruction techniques (e.g. \citet{2001hu, 2003okamoto, 2010smith, 2010bucher, 2011carvalho}). In this paper, we use the CAMB software package \citep{2000lewis} to obtain the numerical lensed and unlensed power spectra  ($C^{TT},  C^{EE}, C^{BB}, C^{TE}$ and $C^{dd}, C^{T d},  C^{E d}$) for our cosmological model with $l \leq 2749$. We then use this predictions to forecast how CMB lensing information will help us constraining our model.

\section{Method\label{Method}}

We apply the Fisher information matrix formalism to a Planck-like experiment \citep{2005bb}, considering both temperature and polarization for the lensend and unlensed CMB spectrum,  and to an experiment such as the future Euclid survey. We forecast the dark energy and the massive neutrinos parameters in our fiducial model. In addition, we check the impact of CMB lensing information on the constraints of the mentioned parameters in a nearly perfect CMB experiment. 

\subsection{Information from CMB}

The Fisher information matrix for the CMB temperature anisotropy and polarization is given by \citep{1997zaldarriaga}

\begin{equation}
\label {fisher_unl}
F_{ij} =\sum_l\sum_{XY}  \frac{\partial C^{X}_l} {\partial p_i} (Cov_l^{-1})_{XY}  \frac{\partial C^{Y}_l} {\partial p_j},
\end{equation}
where $C^{X}_l $ is the power in the $l$th multipole, $X$ stands for $TT$ (temperature), $EE$ (E-mode  polarization), $BB$ (B-mode polarization) and $TE$ (temperature and E-mode polarization cross-correlation). We will not include primordial B-modes in the analysis since the measurement of the primordial $ C^{BB}_l$ by Planck is expected to be noise dominated. Our covariance matrix becomes therefore:

\small
\begin{equation}
\label {cov_array_unl}
Cov_l= \frac {2} {(2l+1)fsky} \left[\begin{array}{rrr}
\Xi_{TTTT} & \Xi_{TTEE} & \Xi_{TTTE}\\
\Xi_{EETT} & \Xi_{EEEE} & \Xi_{EETE}\\
\Xi_{TETT} & \Xi_{TEEE} & \Xi_{TETE}
\end{array}\right].
\end{equation}
\normalsize
Explicit expressions for the matrix elements are given in the appendix Section \ref{Elements of the unlensed covariance matrix}.

For the lensed case we  performed a correction in the covariance matrix elements taking into consideration the power spectrum of the deflection angle and its cross correlation with temperature and E-polarization, $C^{Td}_l$ and  $C^{Ed}_l$.  We used the same procedure introduced in \citep{2006perotto} to obtain the covariance matrix elements using the new information of $C^{Ed}_l$ power spectrum (see appendix Section \ref{Elements of the lensed covariance matrix}).

We also change in this case the unlensed CMB power spectra, $C^{X}_l$, for the lensed ones, $\textrm{\~C}^{X}_l$. Note that in this case we are taking into consideration the B-mode polarization generated by the CMB gravitational lensing from the E-mode polarization. 

When we include these corrections, the covariance matrix becomes:

\small
\begin{equation}
\label {cov_array_lens}
\begin{split}
& Cov_l=  \frac {2} {(2l+1)fsky} \\
&  \left[\begin{array}{rrrrrrr}
\xi_{TTTT} & \xi_{TTEE} & \xi_{TTTE}      & \xi_{TTTd}     & \xi_{TTdd} & \xi_{TTEd}  & 0 \\
\xi_{TTEE} & \xi_{EEEE} & \xi_{EETE}    &  \xi_{EETd}    & \xi_{EEdd} & \xi_{EEEd} &0  \\
\xi_{TTTE} & \xi_{EETE} & \xi_{TETE}     & \xi_{TETd}      &\xi_{TEdd}   & \xi_{TEEd} &0 \\
\xi_{TTTd} & \xi_{EETd}  &\xi_{TETd}     &  \xi_{TdTd}    & \xi_{Tddd}  & \xi_{TdEd}  &0 \\
\xi_{TTdd} & \xi_{EEdd}  &\xi_{TEdd}     &  \xi_{Tddd}    & \xi_{dddd}   & \xi_{ddEd} &0 \\
 \xi_{TTEd} & \xi_{EEEd} &\xi_{TEEd}    &\xi_{TdEd}     &\xi_{ddEd}    & \xi_{EdEd} &0 \\
               0  & 0                    &0                     &0                       &0                   & 0                  &\xi_{BBBB}
\end{array}\right].
\end{split}
\end{equation}
\normalsize

\begin{equation}
\label {cov_l1}
\xi_{TTTT} = \left(\textrm{\~C}^{TT}_l + N^{TT}_l\right)^2\\
\end{equation}

\begin{equation}
\label {cov_l2}
\xi_{EEEE} =  \left(\textrm{\~C}^{EE}_l + N^{PP}_l\right)^2,
\end{equation}

 \begin{equation}
 \label {cov_l3}
\xi_{dddd} =  \left(C^{dd}_l + N^{dd}_l \right)^2,
\end{equation}

\begin{equation}
\label {cov_l4}
\xi_{BBBB} =  \left(\textrm{\~C}^{BB}_l + N^{PP}_l\right)^2,
\end{equation}

\begin{equation}
\label {cov_l5}
\xi_{TETE} =  \frac{1}{2} \left[\left(\textrm{\~C}^{TE}_l \right)^2+ \left(\textrm{\~C}^{TT}_l + N^{TT}_l\right) \left(\textrm{\~C}^{EE}_l + N^{PP}_l\right)\right] ,
\end{equation}

\begin{equation}
\label {cov_l6}
\xi_{TdTd} =  \frac{1}{2} \left[\left(C^{Td}_l \right)^2+ \left(\textrm{\~C}^{TT}_l + N^{TT}_l\right) \left(C^{dd}_l + N^{dd}_l\right)\right] ,
 \end{equation}

\begin{equation}
\label {cov_l7}
\xi_{EdEd} =  \frac{1}{2} \left[\left(C^{Ed}_l \right)^2+ \left(C^{dd}_l + N^{dd}_l\right) \left(\textrm{\~C}^{EE}_l + N^{PP}_l\right)\right] ,
 \end{equation}

 \begin{equation}
 \label {cov_l8}
\xi_{TTEE} =  \left(\textrm{\~C}^{TE}_l \right)^2,
\end{equation}

\begin{equation}
\label {cov_l9}
\xi_{TTdd} = \left(C^{Td}_l \right)^2,
\end{equation}

\begin{equation}
\label {cov_l10}
\xi_{EEdd} =  \left(C^{Ed}_l \right)^2 ,
\end{equation}

\begin{equation}
\label {cov_l11}
\xi_{TEdd} = C^{Ed}_l C^{Td}_l ,
\end{equation}

\begin{equation}
\label {cov_l12}
\xi_{EETd} = C^{Ed}_l C^{TE}_l ,
\end{equation}

\begin{equation}
\label {cov_l19}
\xi_{TTEd} = C^{Td}_l C^{TE}_l ,
\end{equation}

\begin{equation}
\label {cov_l13}
\xi_{TTTE} = \textrm{\~C}^{TE}_l \left(\textrm{\~C}^{TT}_l + N^{TT}_l\right), 
\end{equation}

\begin{equation}
\label {cov_l14}
\xi_{EETE} = \textrm{\~C}^{TE}_l  \left(\textrm{\~C}^{EE}_l + N^{PP}_l\right),
\end{equation}

\begin{equation}
\label {cov_l15}
\xi_{TTTd} = C^{Td}_l \left(\textrm{\~C}^{TT}_l + N^{TT}_l \right)
\end{equation}

\begin{equation}
\label {cov_l16}
\xi_{Tddd} = C^{Td}_l \left(C^{dd}_l + N^{dd}_l \right),
\end{equation}

\begin{equation}
\label {cov_l17}
\xi_{ddEd} = C^{Ed}_l \left(C^{dd}_l + N^{dd}_l\right) ,
\end{equation}

\begin{equation}
\label {cov_l22}
\xi_{EEEd} =  C^{Ed}_l\left(\textrm{\~C}^{EE}_l + N^{PP}_l\right),
\end{equation}

\begin{equation}
\label {cov_l18}
\xi_{TETd} =  \frac{1}{2} \left[C^{Td}_l  \textrm{\~C}^{TE}_l + C^{Ed}_l \left(\textrm{\~C}^{TT}_l+ N^{TT}_l\right) \right] ,
\end{equation}

\begin{equation}
\label {cov_l20}
\xi_{TEEd} =  \frac{1}{2} \left[\left(\textrm{\~C}^{EE}_l + N^{PP}_l\right)  C^{Td}_l  +  C^{Ed}_l \textrm{\~C}^{TE}_l \right] ,
\end{equation}

\begin{equation}
\label {cov_l21}
\xi_{TdEd} =  \frac{1}{2} \left[C^{Ed}_l C^{Td}_l + \left(C^{dd}_l+ N^{dd}_l\right)  \textrm{\~C}^{TE}_l\right] ,
\end{equation}

In these equations,  $N^{TT}_l$ and $N^{PP}_l$ are the gaussian random detector noises for  temperature and polarization respectively, which expression is written using the window function, $B_l^2 = exp[-l(l+1)\theta_{beam}^2 / 8\ln2]$ and the inverse square of the detector noise level for temperature and polarization, $w_T$ and $w_P$.  The Full Width Half Maximum (FWHM), $\theta_{beam}$, is used in radians and $w=(\theta_{beam}\sigma)^{-2}$ is the weight given to each considered  Planck channel  \citep{1999eisenstein} .  We tested two types of experimental setups that can be checked in Tables \ref {planck} and \ref{ideal} .

\small
\begin{equation}
\label {noise1}
N^{TT}_l = [(w_TB_l^{2})_{100}+ (w_TB_l^{2})_{143}+ (w_TB_l^{2})_{217}+ (w_TB_l^{2})_{353}]^{-1}
\end{equation}

\begin{equation}
\label {noise2}
N^{PP}_l = [(w_PB_l^{2})_{100}+(w_PB_l^{2})_{143}+ (w_PB_l^{2})_{217}+  (w_PB_l^{2})_{353}]^{-1}
\end{equation}
\normalsize

Here we used four channels, 100, 143, 217 and 353GHz of the Planck experiment as can be seen from the equations \ref{noise1} and \ref{noise2}.

\begin{table}[h!]
\caption{Planck  \textrm{specifications\footnote{See Planck mission blue book at http://www.rssd.esa.int/SA/PLANCK/docs/Bluebook-ESA-SCI(2005)1\_V2.pdf}.  We used $fsky=0.65$.}\label{planck} }
\begin{ruledtabular}
\begin{tabular}{cccc}
Frequency (GHz) & $\theta_{beam}$ & $\sigma_T(\mu K-arc)$ & $\sigma_P(\mu K-arc)$\\
 100 & 9.5' & 6.82        & 10.9120\\
 143  & 7.1' & 6.0016   & 11.4576\\ 
  217 & 5.0' & 13.0944 & 26.7644\\ 
  353 & 5.0' &40.1016  &81.2944\\
 \end{tabular}
 \end{ruledtabular}
\end{table}  

\begin{table}[h!]
\caption{Nearly perfect experiment suggested by \citet{2003okamoto, 2002hu}). We used $fsky=0.90$.\label{ideal}}
\begin{ruledtabular}
\begin{tabular}{ccc}
 $\theta_{beam}$ & $\sigma_T(\mu K-arc)$ & $\sigma_P(\mu K-arc)$\\
  4.0'  & $ 0.093 \times10^{-6}$   &  $0.13 \times10^{-6}$\\

 \end{tabular}
 \end{ruledtabular}
\end{table}  

\begin{figure*} [h!]
\includegraphics[scale=0.7]{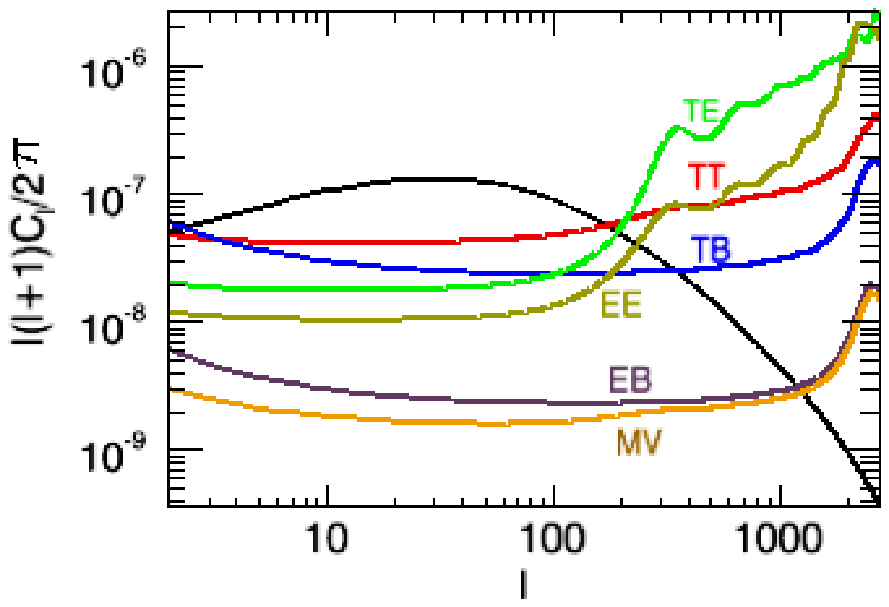}
\includegraphics[scale=0.7]{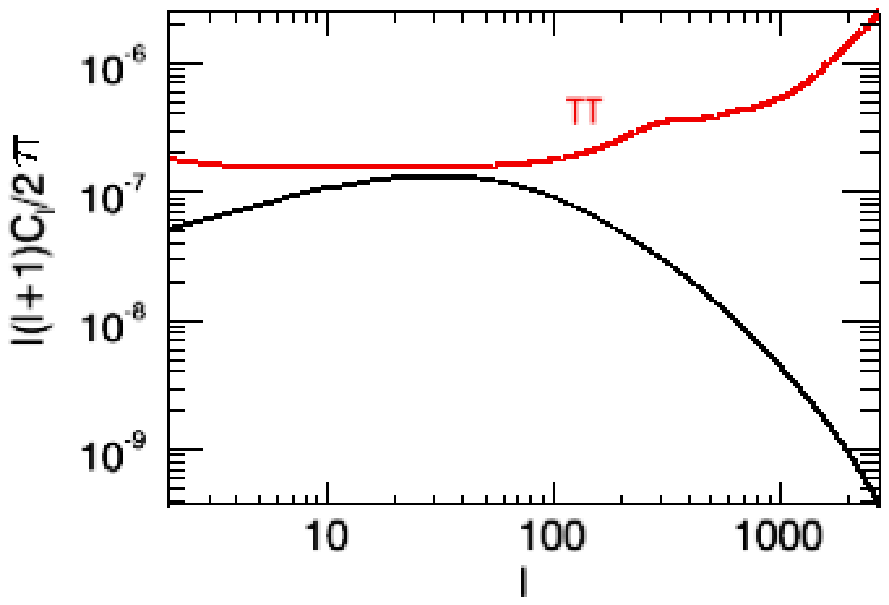}
 
\caption{The CMB deflection field and its quadratic estimator noises for planck experiment (up) specifications and for the nearly ideal experiment (down).} 
\label{quadratic}
\end{figure*}

In addition, $N^{dd}_l$ is the optimal quadratic estimator noise of the deflection field (we consider only the TT quadratic estimator noise for the planck experiment since it provides the best estimator). For the nearly ideal experiment we consider the minimum variance (MV) estimator noise written as a combination of  the noises TB, TT, TE, EE and EB of the quadratic estimators (for a review in the topic, see \citet{2003okamoto, 2002hu})). Figure \ref{quadratic} shows the quadratic estimator noises for our fiducial model considering Planck and the nearly ideal experiment.

%noise of the deflection field and it can be written in the form \citep{2002bhu}:

%\begin{equation}
%\label {noise3}
%N^{dd}_l= \left[ \sum_{l1l2}  \frac{(C_{l2}^{TT}F_{l1ll2} + C_{l1}^{TT}F_{l2ll1})^2}{2(\textrm{\~C}_{l1}^{TT} + N_{l1}^{TT})(\textrm{\~C}_{l2}^{TT} + N_{l2}^{TT})}\right]^{-1} ,
%\end{equation}

%\begin{equation}
%\begin{split}
%F_{l1ll2}= \sqrt{ \frac{(2l_1+1)(2l+1)(2l_2 +1)}{4\pi}}  \left(\begin{array}{rrr}
%l_1 & l & l_2\\
%0 & 0 & _0
%\end{array}\right) \\
%\times  \frac{1}{2} \left[l(l+1) +l_2(l_2+1)-l_1(l_1+1)\right]. \nonumber
%\end{split}
%\end{equation}

\subsection{Information from galaxy survey: Baryonic acoustic oscillation}

We show here how the baryonic acoustic oscillation (BAO) information can be used to forecast errors in the dark energy parameters using the fisher formalism. It was shown by \cite{2003seo} that the Hubble parameter H(z) and the angular diameter distance Da(z) can be measured very precisely by using the BAO information present in the matter power spectrum. H(z) and Da(z) are expected to be determined  as a function of redshift by future galaxy surveys. The goal is then to propagate the errors  on H(z) and Da(z) to the constraints of dark energy parameters. 

We start defining the observed galaxy power spectrum in a reference cosmology (in our case we use the $\Lambda$CDM model), distinguished by the subscript "ref", (different from the true spectrum, referred as no subscript) that will be used to derive the cosmological parameters constraints using a galaxy survey that covers a wide range of redshifts.  Following \cite{2003seo},

\begin{equation}
\label {p_obs}
P_{obs}(k_{ref \perp}, k_{ref \parallel}) = \frac{Da(z)^2_{ref } \times H(z)}{Da(z)^2 \times H(z)_{ref }}P_{g}(k_{ref}, k_{ref}) + P_{shot}
\end{equation}

Where the Hubble parameter H(z) in a flat Universe is related to the dark energy parameters through

\small
\begin{equation}
\label {hubble_parameter}
H(z) = H_0 \sqrt{\Omega_m(1+z)^3 + \Omega_de(1+z)^{3(1+w_0+w_a)}\exp{(3w_a(a-1))}},
\end{equation}
\normalsize

and the angular diameter distance is defined as

\begin{equation}
\label{da2}
D_a(z)= \frac{c}{1+z}\int_{0}^{z} \frac{dz}{H(z)},
\end{equation}. 

$P_{shot}$ is the unknown Poisson shot noise.

The wavenumbers across and along the line of sight are denoted by $k_{\perp}$ and $k_{\parallel}$.  It is important to point out that the wavenumbers in the reference cosmology are related to the ones in the true cosmology by

\begin{equation}
\label {k_ref}
\begin{split}
 k_{ref \perp} =  k_{\perp} \frac{Da(z)}{Da(z)_{ref}} \\
 k_{ref \parallel} =  k_{\parallel} \frac{H(z)_{ref}}{H(z)}.
\end{split}
\end{equation}

Moreover, we define the galaxy power spectrum, $P_g$ , including the redshift distortions: 

\small
\begin{equation}
\label {p_g}
\begin{split}
P_{g}(k_{ref \perp}, k_{ref \parallel}) =b^2(z)\left( 1+ \beta \mu^2 \right)^2 \left(\frac{G(z)}{G(0)}\right)^2 \\
\times P_{matter,z=0}(k) e^ {-k^2 \mu^2 \sigma_r^2},
\end{split}
\end{equation}
\normalsize

where $\mu = $ {\bf{k}} $\cdot$ {\bf{\^r}}/k , being {\bf {\^r}} the unit vector along the line of sight and the linear matter power spectrum, $P_{matter,z=0}(k)$, was generated using CAMB software package \citep{2000lewis} and COBE normalized. The $k_{max}$ is chosen in a way to exclude information from non linear regime where equation \ref{p_obs} is inaccurate (see \cite{2003seo}). For an approach considering the non-linear regime see, for example, \cite{2005seo,2007seo, 2010wang,  2012wang}. Moreover, for the impact of precisely modeling systematic effects, such as the non-linear clustering and redshift space distortions, in the evolution of BAO  see \cite{2008crocce, 2009taruya, 2010taruya}. The exponential damping factor is due to redshift uncertainties, where $\sigma_r=c\sigma_z/H(z)$.  $G(z)$, $\beta(z)$ and $b(z)$ are the growth function,  the linear redshift space distortion parameter and the linear galaxy bias respectively. We use a growth factor dependent on the dark energy parameter and massive neutrinos effect computed by \cite{2008kiakotou}. The growth rate of perturbations is defined as 

\begin{equation}
\label {gz}
f \equiv \frac{d\ln G}{d \ln a},
\end{equation}

where the growth function G(z) is related to the density of matter. In a matter dominated Universe $f \approx \Omega_m(z)^{0.6}$, with $\Omega_m(z)= H_0^2\Omega_m(1+z)^3/H^2(z)$.  More generally, we use $f= \nu \Omega_m(z)^{\alpha}$ with

\begin{equation}
\label {alpha}
\begin{split}
\alpha = \alpha_0 + \alpha_1 [1-\Omega_m(z)], \\
\alpha_0=\frac{3}{5- \frac{w}{1-w}},\\
\alpha_1=\frac{3}{125} \frac{(1-w)(1-3w/2)}{(1-6w/5)^3}.
\end{split}
\end{equation}

$\nu$ is  a numerical function dependent on $\Omega_{de}$ and $f_\nu=\Omega_\nu/\Omega_m$  (see Equation 17 and Table 5 of \cite{2008kiakotou}).

The linear redshift space distortion is also defined as a function os the growth rate and the galaxy bias.

\begin{equation}
\label {beta}
\beta(z)  \equiv \frac{f}{b(z)}.
\end{equation}

\subsubsection{Fisher formalism}

The Fisher information matrix for the matter power spectrum obtained from galaxy surveys is given by \citep{1997tegmark}

\small
\begin{equation}
\label {fisher_gal2}
F_{ij} =\int_{-1}^1\int_{k_{\textrm{min}}}^{k_{\textrm{max}}} \frac{\partial \ln P(k, \mu)} {\partial p_i}  \frac{\partial \ln P(k,\mu)} {\partial p_j}V_{\textrm{eff}}(k,\mu) \frac{2\pi k^2 dkd\mu} {2(2\pi)^3}.
\end{equation}
\normalsize

The effective volume of the survey for a constant comoving number density  is given by

\begin{equation}
\label {veff}
V_{\textrm{eff}}(k,\mu) = \left[ \frac{\bar{n}(r) P_g(k,\mu)} {1+\bar{n}(r)P_g(k,\mu)} \right]^2 V{survey} .
\end{equation}

We use information of an Euclid like survey  with area of 20000 $deg^2$, redshift accuracy of  $\sigma_z/(1+z)=0.001$ and a redshift range $0.5 \leq z\leq 2.1$. Finally we dived our forecast into 15 redshift slices of $\Delta z =0.1$ centered in $z_i$. We chose the  initial set of parameters $P=\{ h^2\Omega_b, h^2\Omega_c, h^2\Omega_\nu, H(z_i), Da(z_i), G(z_i), \beta(z_i), P^i_{shot} \}$. For each redshift bin we use the specifications on Table \ref{survey} (see \cite{2012pavlov, 2012diporto} and references therein).

\begin{table}[h!]
\begin{center}
\caption{Values of $k_{max}$, the galaxy bias and the galaxy density for each redshift bin.}
\label{survey}
\begin{tabular}{cccc}
\hline\hline
$ z_i $& $K_{max}$ ($hMpc^{-1}$) & b(z) & n(z) $\times 10^{-3} (h/ Mpc)^3$\\
\hline
 0.55 & 0.144& 1.0423       & 3.56\\
 0.65  & 0.153& 1.0668      & 3.56\\ 
  0.75 & 0.163 & 1.1084     & 2.42\\ 
  0.85 & 0.174& 1.1145      &2.42\\
  0.95 &0.185 & 1.1107      &1.81\\
  1.05 &0.197 & 1.1652       &1.81\\
  1.15 & 0.2    &  1.2262       &1.44\\
  1.25 & 0.2    &   1.2769       &1.44\\
  1.35 &0.2     &   1.2960       &0.99\\
  1.45 &0.2     &  1.3159        &0.99\\
  1.55 &0.2     &   1.4416        &0.55\\
  1.65 &0.2     &   1.4915        &0.55\\
  1.75 &0.2     &   1.4973       & 0.29\\
  1.85 &0.2     &   1.5332        &0.29\\
  1.95 &0.2     &   1.5705      &0.15\\
\hline\hline
 \end{tabular}
\end{center}
\end{table}  

To obtain the constraints on our final set of parameters $Q=\{ h^2\Omega_b, h^2\Omega_c, h^2\Omega_\nu, w_0, w_a \}$, first we marginalize our first fisher matrix over $G(z_i), \beta(z_i), P^i_{shot}$ and use this sub matrix to change into the desired variables as

\begin{equation}
\label {fisher_gal3}
F_{DE,ij} =\sum_{\alpha, \beta} \frac{\partial P_\alpha}{\partial Q_i} F^{sub}_{\alpha \beta}\frac{\partial P_\beta}{\partial Q_j}
\end{equation}

\subsection{Information from galaxy survey: Weak lensing}

In this subsection, we show how weak lensing (WL) information can improve the constraints on cosmological parameters, in our case specially for dark energy  and neutrinos' densities parameters using the fisher formalism. The observable, in weak lensing surveys, is the convergence power spectrum.  In the analysis presented in this paper, we use  an extension of the CAMB software with Halofit approximation (recently updated according to \cite{2012takahashi}) to generate the convergence power spectra  $P_{ij}$, where the subscripts $i$ and $j$ stand to the lensed galaxies redshift bins. The fisher matrix for weak lensing is then given by \citep{1999hu}

\small
\begin{equation}
\label {fisher_wl}
F_{\alpha \beta} =f_{\textrm{sky}}\sum_{l} \frac{(2l+1)} {2} \frac{\partial P_{ij}} {\partial p_{\alpha}}(C^{-1})_{jk} \frac{\partial P_{km}} {\partial p_{\beta}}(C^{-1})_{mi}.
\end{equation}
\normalsize

The covariance matrix is defined as

\begin{equation}
\label {cov_wl}
C_{jk}= P_{jk} + \delta_{jk} \langle \gamma^2_{\textrm {int}}  \rangle n_j^{-1},
\end{equation}

being $\gamma_{\textrm {int}}$ the r.m.s intrinsic shear and $n_j$ the number of galaxies per steradian in the j-th bin

\begin{equation}
\label {n_wl}
n_{j}= 3600d  \left(\frac{180} {\pi}\right)^2 {\textrm {\^n}}_j.
\end{equation}

In the equation above, d is the number of galaxies per square arcminute and ${\textrm {\^n}}_j$ is fraction of sources belonging to a certain bin. We compute our calculations considering an Euclid-like experiment following \cite{2008bb} with $f_{\textrm{sky}}=1/2$, $d=40$ and  $\langle \gamma^2_{\textrm {int}}  \rangle^{1/2} = 0.22$ .  We take the range $0.5 \leq z \leq 2.0$ and divide it into 4 equal-galaxy-number  bins. We also consider $10 \leq l \leq 10000$. In this case, we tested the analysis for a maximum multipole of 3000 and no significant change was found confirming that both larger and smaller multipoles do not give a significant contribution to the results (see  \cite{2009hollenstein}). The photo-z error is assumed to be normal distributed with variance $\sigma_z=0.005(1+z)$. It is important to point out that non-gaussian errors can be significant in the measurements of weak lensing, degrading the signal-to-noise ratio of convergence power spectrum \citep{2009sato} and therefore the marginalized errors on individual parameters by a few percent \citep{2009btakada}.

%\section{Fiducial models}
 %\label{Fiducial models}

%We assume cosmologies with all of them having as fiducial values $w_0=-0.95$ and $w_a=0$. The first model is assumed to have 3 massive neutrinos, being $\Omega_m =\Omega_c + \Omega_b + \Omega_\nu$. As a second model we assume two massive neutrinos with identical mass and one massless in a inverted hierarchy mass splitting. In the third model, we will assume two massless and one massive neutrino in a normal hierarchy mass splitting.   The free cosmological parameters for each model assume the values in.

%\begin{table}[h!]
%\begin{center}
%\caption{Fiducial models}
%\label{models}
%\begin{tabular}{cccc}
%\hline\hline
%Parameter & Model 1  & Model 2 &  Model 3 \\
%\hline
 %$h^2 \Omega_b$       & 0.02219       &0.02219       & 0.02258     \\
 %$h^2 \Omega_c$       & 0.01122        &0.1122         & 0.01109    \\ 
  %$h^2 \Omega_\nu$  & 0.01               & 0.0027         & 0.0021       \\ 
  %$M_{\nu1}$                & 0.310 eV       & 0                  & 0                  \\
  %$M_{\nu2}$               &  0.310 eV       & 0.127eV     & 0                  \\
  %$M_{\nu3}$             &  0.310 eV        &  0.127 eV      & 0.198 eV    \\
  
%\hline\hline
 %\end{tabular}
%\end{center}
%\end{table}  

\section{Results\label{Results}}

We performed the forecast for Planck alone, with and without considering CMB lensing. Moreover, we introduced the Euclid forecast, combining the results approximately  as 

\begin{equation}
\label {fisher_unl_len_euclid}
F^{\textrm{Total}}_{ij} = F^{\textrm{Planck}}_{ij} + F^{\textrm{Euclid (BAO)}}_{ij} + F^{\textrm{Euclid (WL)}}_{ij}.
\end{equation}

It was shown by \citet{2009hollenstein}  that the covariance between the measurements of cosmic shear tomography and the CMB lensing can be safely neglected since the redshifts in which they are probed are quite distinct from each other.

%We considered the three previously mentioned models (Section \ref{Fiducial models}). In Table \ref{tbl-modelo1} and Figure \ref{fisher_modelo1}, we show the 1 sigma errors and the 2 sigma fisher contours for model 1. As the best upper limits for the neutrino density, $w_0$ and $w_a$ we  found, for the combined Planck + Euclid, $0.0097 < h^2\Omega_{\nu} < 0.0103 $, $ -0.954< w_0 < -0.946$ and $-0.03< w_a < 0.03$ (all of them with 2 sigma confidence level).  We can roughly compare our results with \cite{2011carboneb}  since they are not using the flat sky approximation. They found that the 1 sigma errors on $w_0$ and $w_a$ are 0.0732 and 0.176 respectively (using Planck + Euclid (BAO) + BOSS) against 0.0020 and 0.015 (using Planck (+ CMB lensing) + Euclid (BAO + WL))  found by us. It is important to point out that our better result is also related to the fact that we use CMB information directly to constraint both $w_0$ and $w_a$. \cite{2011carboneb}  do not use CMB to constraint directly these parameters, but only to constraint other parameters of interest that are then summed up with the Euclid survey forecast that include $w_0$ and $w_a$.

\begin{figure*} [h!]
\includegraphics[scale=0.5]{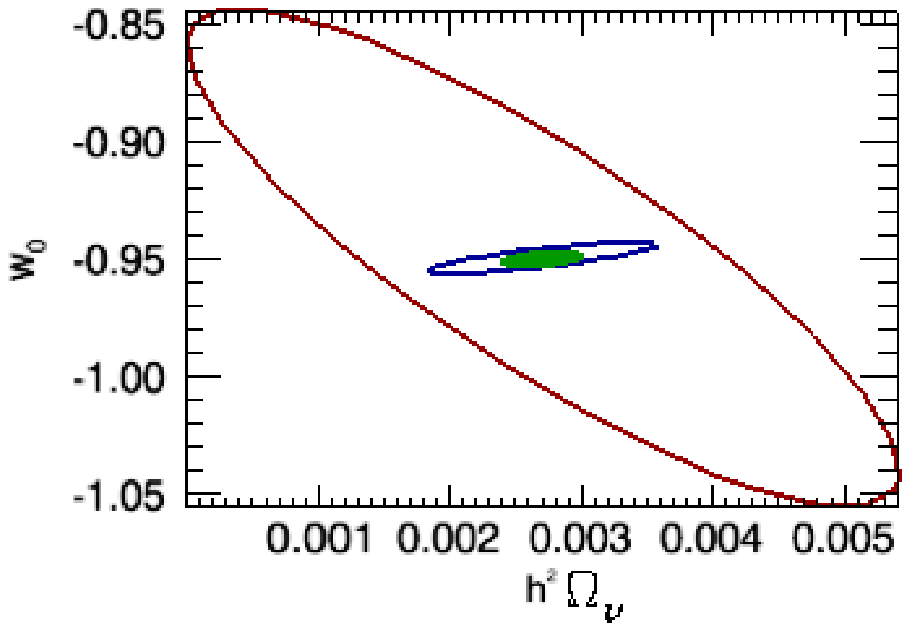}
\includegraphics[scale=0.5]{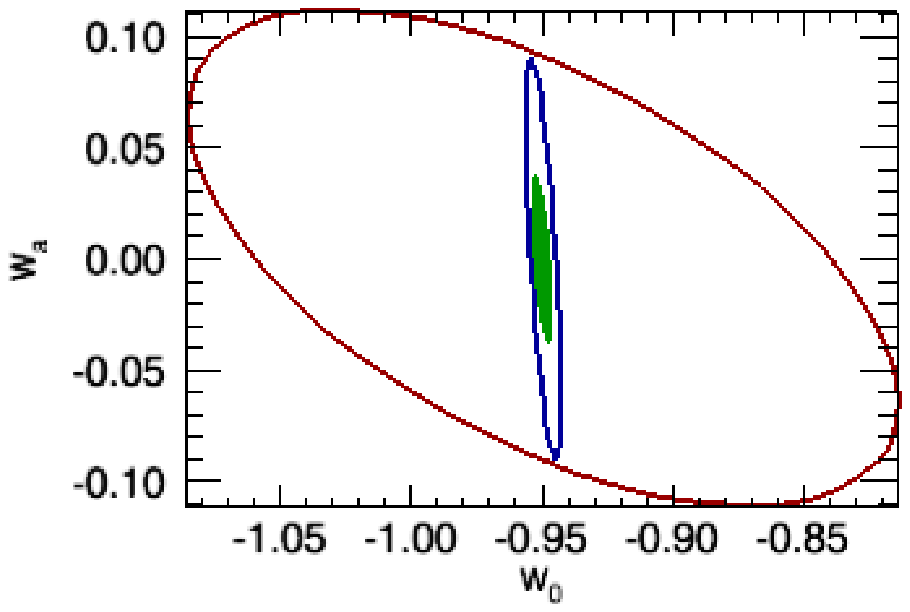}
\includegraphics[scale=0.5]{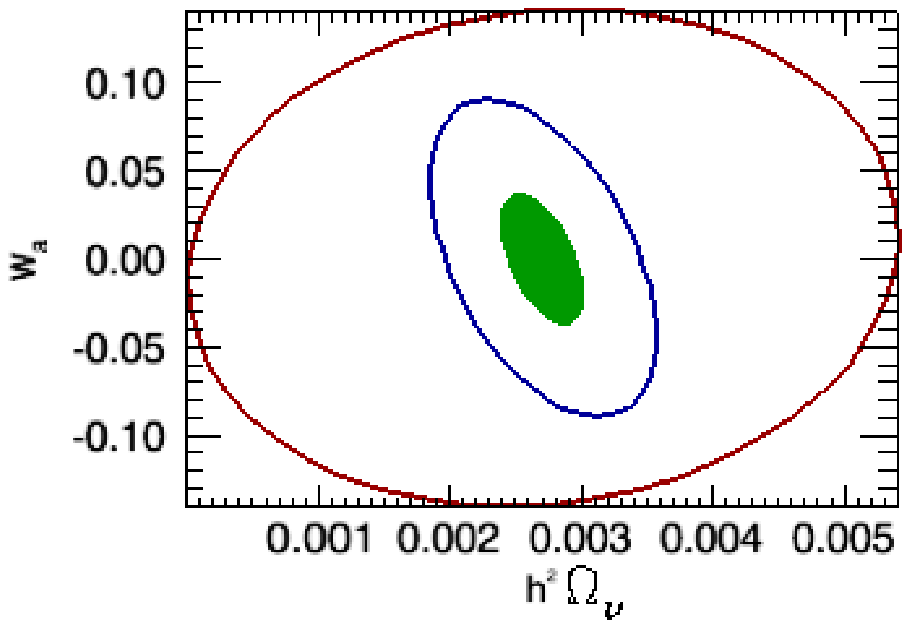}

\caption{Fisher contours for our fiducial model. The  contours represent 95.4\% C.L.  for  CMB (red), for Euclid galaxy survey (blue) and for the combination of CMB + Euclid (filled green) (see Table \ref {tbl-modelo2}). } 
\label{fisher_modelo2}
\end{figure*}

We found the best limits for the neutrino density, $w_0$ and $w_a$ in the combined Planck (with lensing)+ Euclid (BAO +WL). We have that $0.00244 < h^2\Omega_{\nu} < 0.00296 $, $ -0.953< w_0 < -0.947$ and $-0.03< w_a < 0.03$ (95\% C.L) as it can be inferred from column 6 of Table \ref{tbl-modelo2} .  The Figure of Merit (FoM), described as the reciprocal of the 95\% confidence limit's area of the error ellipse from the plane $w_0$ - $w_a$ \citep{2006albrecht},  of Euclid (BAO +WL) is a factor $ \sim$ 29  higher than the FoM for Planck alone even when CMB lensing is considered. Euclid will be able to strongly constraint the late-universe parameters.  The combined result Planck (with lensing)+ Euclid (BAO +WL) still improves the FoM in respect to Euclid (BAO +WL) in a factor of $\sim$ 3. In Figure \ref{fisher_modelo2} we show the 2 sigma fisher contours. 

%For model 2, we have that $0.00244 < h^2\Omega_{\nu} < 0.00296 $, $ -0.953< w_0 < -0.947$ and $-0.03< w_a < 0.03$ (95\% C.L) as it can be inferred from Table \ref{tbl-modelo2} (see Figure \ref{fisher_modelo2} for the fisher constraints). Finally in Table \ref{tbl-modelo3} and Figure \ref{fisher_modelo3}, we show the 1 sigma errors and the fisher ellipses for model 3, getting with 95\% C.L.  $0.001904< h^2\Omega_{\nu} < 0.002296 $, $ -0.9526< w_0 < -0.9474$ and $-0.022< w_a < 0.022$.

\begin{figure*} [h!]
\includegraphics[scale=0.5]{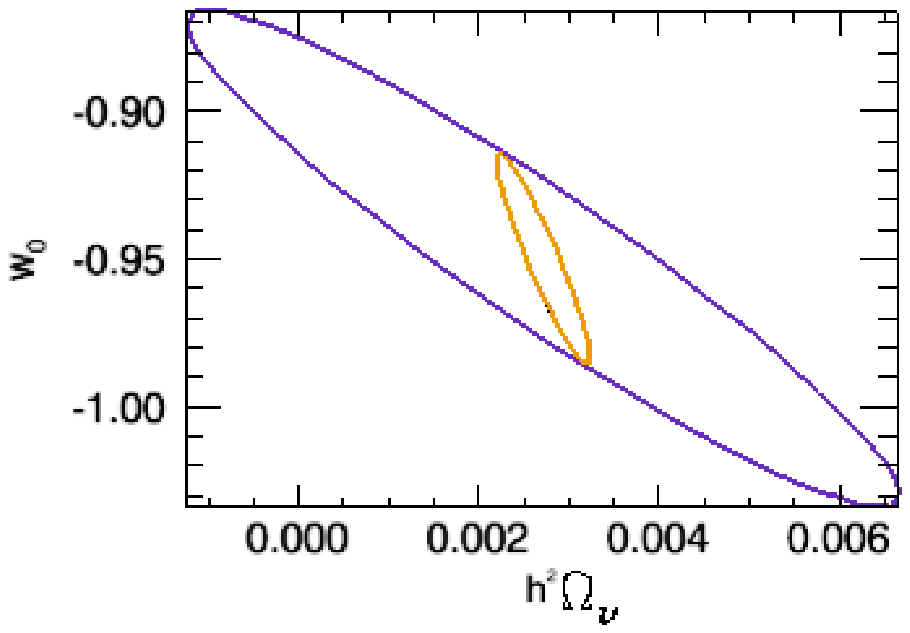}
\includegraphics[scale=0.5]{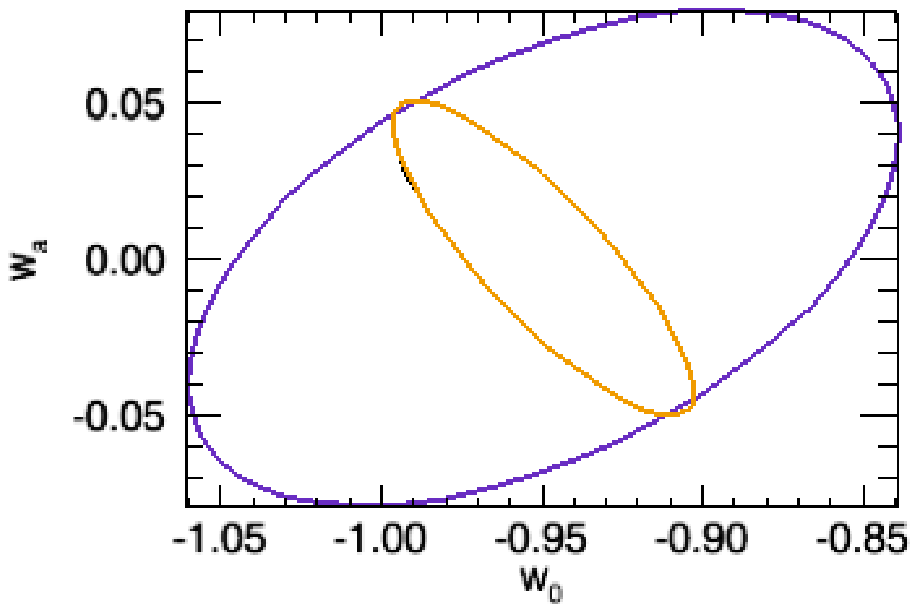} \\
\includegraphics[scale=0.5]{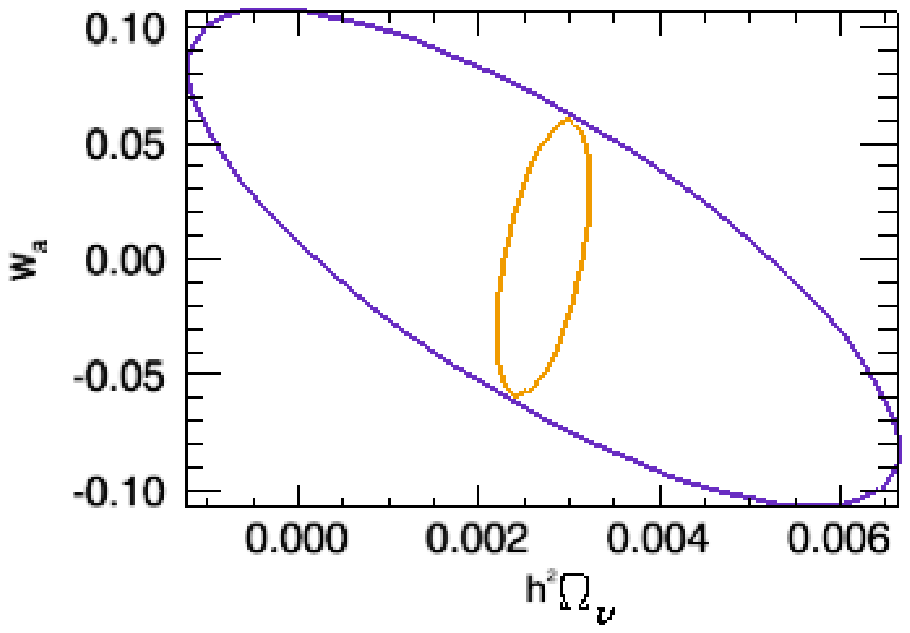} 

\caption{Fisher contours with and without lensing information in orange and purple respectively. The  contours represent 95.4\% C.L.  for  CMB.  (see Table \ref {tbl-modelo2_ideal}). } 
\label{fisher_modelo2_ideal}
\end{figure*}

 In table \ref{tbl-modelo2_ideal} we show how CMB lensing could affect the parameters constraints for an almost ideal experiment. In this case, we see a substantial improvement of about 6 times in the FoM when CMB lensing is considered. Comparing also with Planck experiment, the nearly ideal experiment improves the FoM by a factor of 5 without considering lensing information in any case. For the neutrino parameter we have that $ h^2\Omega_{\nu} < 0.005872$ without lensing and $ h^2\Omega_{\nu} < 0.003113$ when lensing CMB is considered. On the other hand, the use of the cross power spectrum between the deflection angle and the E mode polarization makes no significant impact on any cosmological parameter constraint. Figure \ref{fisher_modelo2_ideal} shows the fisher contours for the unlensed and lensed analysis. Note that the 2 sigma contour obtained when we  use $C^{dd}$ and $C^{td}$ power spectra overlap the 2 sigma contour when we also add the $C^{ed}$ power spectrum.

\begin{table*}
\begin{center}
\caption{Marginalized errors for $\Lambda$CDM model with two massive neutrinos with identical mass and one massless in a inverted hierarchy mass splitting ($m_{\nu}$ = 0.125 eV) .\label{tbl-modelo2} }
\begin{tabular}{cccccc}
\hline
Parameter & Fiducial            & Planck         & Planck          & EUCLID                           & Planck + EUCLID     \\
                    &value                  & T + P            & T+ P+ lens  &  (BAO + WL)                    &                                      \\
 \hline \hline

 $h^2 \Omega_b$       & 0.02219        &0.00012     & 0.00012     &0.00034    & 7.9e-05       \\ 
 $h^2 \Omega_c$        &0.01122        & 0.00080    & 0.00070     &0.00011    & 8.2e-05    \\ 
 $h^2 \Omega_{\nu}$ & 0.0027          & 0.0036     & 0.0011       & 0.00035      & 0.00013  \\
 $w_0$                          & -0.95             &0.084        & 0.042           & 0.0027         & 0.0015     \\
 $w_a$                          & 0                    &0.084       & 0.057          &0.036              & 0.015       \\
 FoM                              &-                       & 8.21         &25.81           &732.97            &2909.13  \\
Relative FoM \footnote{Relative FoM in respecto to Planck (T+P) without CMB lensing.}&- & 1& 3.14& 89.3  & 354.34   \\
 \hline
 \end{tabular} 
\end{center}
\end{table*}

\begin{table*}
\begin{center}
\caption{Marginalized errors for $\Lambda$CDM model with two massive neutrinos with identical mass and one massless in a inverted hierarchy mass splitting ($m_{\nu}$ = 0.125 eV) for a nearly perfect experiment .\label{tbl-modelo2_ideal} }
\begin{tabular}{ccccc}
\hline
Parameter & Fiducial     &                                  & T + P+                                             &  T+ P  \\      
                    &value          &   T+P (unlensed)   & lens ($C^{dd}$ and $C^{td})$    &lens ($C^{dd} $, $ C^{td}$ and $C^{ed}$)  \\
 \hline \hline

 $h^2 \Omega_b$       & 0.02219 & 2.4778e-05        &2.2296e-05    & 2.2285e-05  \\ 
 $h^2 \Omega_c$        &0.01122 &0.0004420           &0.0003073    & 0.0003071  \\ 
 $h^2 \Omega_{\nu}$ & 0.0027   &0.0015860           &0.0002065    & 0.0002064 \\
 $w_0$                          & -0.95      &0.0338432           &0.0142514     & 0.0142416    \\
 $w_a$                          & 0             & 0.0338432          &0.0238881      & 0.0238811  \\
 FoM                              &-               & 41.83                    & 255.43             & 255.63    \\
 Relative FoM              &-               & 1                            &6.106                 &6.111     \\
 \hline
 \end{tabular} 
\end{center}
\end{table*}

%\begin{figure*} 
%\includegraphics[scale=0.5]{./figures/Fig3a.eps}
%\includegraphics[scale=0.5]{./figures/Fig3b.eps} 
%\includegraphics[scale=0.5]{./figures/Fig3c.eps} 

%\caption{Fisher contours for our model 3. The  contours represent 95.4\% C.L.  for  CMB (red), for Euclid galaxy survey (blue) and for the combination of CMB + Euclid (filled green) (see Table \ref {tbl-modelo3}).}  
%\label{fisher_modelo3}  
%\end{figure*}

\section{Discussion and conclusions\label{Discussion and conclusions}}

We are considering massive neutrinos and a time evolving equation of state in the $\Lambda$CDM model. Using the Fisher formalism, we obtained the best constraints possible for  $h^2\Omega_{\nu} $, $w_0$ and $ w_a$ considering Planck and Euclid survey.

One of the goals of this work has been to quantify the influence of CMB lensing information in the constraints of the parameters of interest, especially  $h^2\Omega_{\nu} $, $w_0$ and $ w_a$.  We saw on columns 3 and 4  from Table \ref{tbl-modelo2} that we improve the constraints in basically all the studied cosmological parameters. The  improvement  found on $h^2\Omega_{\nu} $ for the Planck 1 sigma error  alone varied from approximately 133\% to  40\% from its fiducial value  without using CMB lensing and using CMB lensing information respectively.  For $w_0$  the error is 4.4\% smaller when CMB lensing is taken into consideration. The 2 sigma constraint on $w_a$ from Planck alone varies from $-0.168< w_a < 0.168$ (without CMB lensing) and $-0.114< w_a < 0.114$.

When we add Euclid information to Planck information, we get an impressive improvement of 128.2\% on the 1 sigma error of $h^2\Omega_{\nu}$  considering the results from Planck without lensing and Planck (+ lensing) + Euclid (BAO + WL).  An improvement of approximately 9\% in the error of $w_0$ was found when including Euclid information to the Planck forecast. 

In the case of a nearly perfect CMB experiment, as mentioned before, CMB lensing improved all the constraints of the tested parameters. It is clear from the analysis that CMB lensing can play an important role in constraining cosmological parameters in future CMB experiments and must be taken into account. On another hand,  the $C^{ed}$ power spectra can be safely neglected in near future CMB experiments since its contribution to the parameters constraints is minimum. 

\acknowledgements
The authors would like to thank Carmelita Carbone and Yun Wang for useful discussions. This work has been partially supported by ASI contract for Planck LFI activity of Phase E2.

\appendix

\section{Elements of the unlensed covariance matrix}\label{Elements of the unlensed covariance matrix}

The elements of the covariance matrix in the unlensed case are:

\begin{equation}
\label {cov1}
\Xi_{TTTT} = (C^{TT}_l + N^{TT}_l)^2,
\end{equation}

\begin{equation}
\label {cov2}
\Xi_{EEEE} =  (C^{EE}_l + N^{PP}_l)^2,
\end{equation}

\begin{equation}
\label {cov3}
\Xi_{BBBB} =  (C^{BB}_l + N^{PP}_l)^2,
\end{equation}

\begin{equation}
\label {cov4}
\begin{split}
\Xi_{TETE} =& (C^{TE}_l )^2+ (C^{TT}_l + N^{TT}_l) \\
 &\times (C^{EE}_l + N^{PP}_l),
\end{split}
\end{equation}

 \begin{equation}
 \label {cov5}
\Xi_{TTEE} =  (C^{TE}_l )^2,
\end{equation}

\begin{equation}
\label {cov6}
\Xi_{TTTE} = C^{TE}_l  (C^{TT}_l + N^{TT}_l),
\end{equation}

\begin{equation}
\label {cov7}
\Xi_{EETE} = C^{TE}_l  (C^{EE}_l + N^{PP}_l),
\end{equation}

\begin{equation}
\label {cov8}
\Xi_{TTBB} = \Xi_{EEBB} = \Xi_{TEBB} = 0.
\end{equation}

\section{Elements of the lensed covariance matrix}\label{Elements of the lensed covariance matrix}

First of all, we make use of the effective $\chi^2$ defined in Equation (3.3) of \cite{2006perotto}.

\begin{equation}
\label {chi_squared}
\chi^2_{eff} = \sum_l (2l+1) \left (\frac{D}{|C|} + \ln \frac{|C|}{|\textrm{\^C}|} -3\right)
\end{equation}

where is our case D is defined to be

\small
\begin{equation}
\label {d_equation}
\begin{split}
D= \textrm{\^C}^{TT}C^{EE}C^{dd}C^{BB} + C^{TT} \textrm{\^C}^{EE}C^{dd}C^{BB}+ C^{TT} C^{EE} \textrm{\^C}^{dd}C^{BB}\\
+ C^{TT} C^{EE} C^{dd}\textrm{\^C}^{BB} + 2(\textrm{\^C}^{TE}C^{Ed}C^{Td}C^{BB} + C^{TE}\textrm{\^C}^{Ed}C^{Td}C^{BB}\\
 + C^{TE}C^{Ed}\textrm{\^C}^{Td}C^{BB} + C^{TE}C^{Ed}C^{Td}\textrm{\^C}^{BB})- C^{Ed}(\textrm{\^C}^{TT}C^{BB}C^{Ed}\\
 +C^{TT}\textrm{\^C}^{BB}C^{Ed}+2C^{TT}C^{BB}\textrm{\^C}^{Ed})  - C^{TE}(\textrm{\^C}^{dd}C^{BB}C^{TE}\\
 +C^{dd}\textrm{\^C}^{BB}C^{TE} +2C^{dd}C^{BB}\textrm{\^C}^{TE})  - C^{Td}(\textrm{\^C}^{EE}C^{BB}C^{Td}\\+C^{EE}\textrm{\^C}^{BB}C^{Td} +2C^{EE}C^{BB}\textrm{\^C}^{Td},
\end{split}
\end{equation}
\normalsize

and $|C|$ , $|\textrm{\^C}|$ are the determinants of the theoretical and observed data covariance matrices 

\small
\begin{equation}
\label {obs_determintant}
\begin{split}
|\textrm{\^C}|= \textrm{\^C}^{TT}\textrm{\^C}^{EE}\textrm{\^C}^{dd}\textrm{\^C}^{BB} + 2\textrm{\^C}^{TE}\textrm{\^C}^{Ed}\textrm{\^C}^{Td}\textrm{\^C}^{BB} \\
-\textrm{\^C}^{TT}\textrm{\^C}^{BB}(\textrm{\^C}^{Ed})^2-\textrm{\^C}^{dd}\textrm{\^C}^{BB}(\textrm{\^C}^{TE})^2 -\textrm{\^C}^{EE}\textrm{\^C}^{BB}(\textrm{\^C}^{Td})^2,
\end{split}
\end{equation}
\normalsize

\small
\begin{equation}
\label {theoretical_determintant}
\begin{split}
|C|= C^{TT}C^{EE}C^{dd}C^{BB} + 2C^{TE}C^{Ed}C^{Td}C^{BB} \\
-C^{TT}C^{BB}(C^{Ed})^2-C^{dd}C^{BB}(C^{TE})^2 -C^{EE}C^{BB}(C^{Td})^2.
\end{split}
\end{equation}
\normalsize

The theoretical covariance matrix M is given by
\small
\begin{equation}
\label {cov_ed}
M=\left[\begin{array}{rrrr}
C^{TT} & C^{TE} & C^{Td}      & 0     \\
C^{TE} & C^{EE} & C^{Ed}     &  0    \\
C^{Td} & C^{Ed} & C^{dd}     & 0     \\
0           & 0           &0                 & C^{BB}.
\end{array}\right].
\end{equation}
\normalsize

The fisher matrix  information is  then derived from the second order derivative of the likelihood function, $L$, from an observing data set $x$ given the real parameters $p_1,p_2,p_3,...,p_n$:

\begin{equation}
\label {likelihood}
F_{ij}=-\left \langle \frac {\partial^2 \ln L} {\partial p_i \partial p_j} \right \rangle_x,
\end{equation}

knowing that $\chi^2_{eff} \equiv -2\ln L$, we derived the new  elements for the covariance matrix \ref{cov_array_lens}.

\end{document}